\documentclass[aps,prl,twocolumn,floats,showpacs,superscriptaddress]{revtex4-1}
\usepackage{graphicx,epsfig}% Include figure files
\usepackage{times}
\usepackage{graphics,dcolumn,bm,float}
\usepackage{amssymb,amsmath,rotate,color}

\usepackage{graphicx}

\begin{document}
\unitlength 1 cm
\newcommand{\be}{\begin{equation}}
\newcommand{\ee}{\end{equation}}
\newcommand{\bearr}{\begin{eqnarray}}
\newcommand{\eearr}{\end{eqnarray}}
\newcommand{\nn}{\nonumber}
\newcommand{\la}{\langle}
\newcommand{\ra}{\rangle}
\newcommand{\cd}{c^\dagger}
\newcommand{\vd}{v^\dagger}
\newcommand{\ad}{a^\dagger}
\newcommand{\bd}{b^\dagger}
\newcommand{\kt}{{\tilde{k}}}
\newcommand{\pt}{{\tilde{p}}}
\newcommand{\qt}{{\tilde{q}}}
\newcommand{\eps}{\varepsilon}
\newcommand{\vk}{{\mathbf k}}
\newcommand{\vp}{{\mathbf p}}
\newcommand{\vq}{{\mathbf q}}
\newcommand{\vK}{{\mathbf K}}
\newcommand{\vP}{{\mathbf P}}
\newcommand{\vQ}{{\mathbf Q}}
\newcommand{\vkp}{\mathbf {k'}}
\newcommand{\vpp}{\mathbf {p'}}
\newcommand{\vqp}{\mathbf {q'}}
\newcommand{\bk}{{\mathbf k}}
\newcommand{\bp}{{\mathbf p}}
\newcommand{\bq}{{\mathbf q}}
\newcommand{\br}{{\mathbf r}}
\newcommand{\bR}{{\mathbf R}}
\newcommand{\up}{\uparrow}
\newcommand{\down}{\downarrow}
\newcommand{\fns}{\footnotesize}
\newcommand{\ns}{\normalsize}
\newcommand{\cdag}{c^{\dagger}}

\title{Gapless chiral excitons in thin films of topological insulators}

\author{S. A. Jafari{\footnote {Electronic address: jafari@physics.sharif.edu}}}
\affiliation{Department of Physics, Sharif University of Technology, Tehran 11155-9161, Iran}
\affiliation{Center of excellence for Complex Systems and Condensed Matter (CSCM), Sharif University of Technology, Tehran 1458889694, Iran}
\affiliation{School of Physics, Institute for Research in Fundamental Sciences, Tehran 19395-5531, Iran}

%\author{G. Baskaran}
%\affiliation{Institute of Mathematical Sciences, Chennai 600113, India}

\begin{abstract}
In a nanoscopic thin film of a strong topological insulator (TI)  
the Coulomb interaction in the channel that exchanges the two electrons
with the same chirality in two different planes of the slab takes advantage of 
the minus sign resulting from such "exchange" and gives rise to a bound
state between the positive energy states in one surface and the negative energy 
states in the opposite surfaces. Therefore particle and hole pairs in the {\em undoped}
Dirac cone of the TI thin film form an inter-surface spin-singlet state
that lies below the continuum of free particle-hole pairs. 
This mode is similar to the excitons of semiconductors, albeit
formed between the electron and hole pairs from two different
two-dimensional surfaces. 
For low-momenta the dispersion relation characterizing this 
collective mode is linear. 
Experimental comparison of two slabs with different thicknesses can capture the
exponential dependence of the present effect on the slab thickness.
\end{abstract}
\maketitle

\section{Introduction}
Strong topological insulators (TI) are characterized by a bulk gap,
while the surface surrounding the material hosts odd number of gapless
Dirac fermions~\cite{zhang2009,hasan}.
The classification of the topological insulators are based on
the non-interacting picture and their basic band structure~\cite{Shen}.
%This leaves a room for the investigation of 
%the effects of interactions of various forms on the 
%basic properties of TIs, including the nature of their
%collective excitations and the origin of possible instabilities arising
%from the interactions.
The gapless helical surface states are characteristic of topological
insulators as long as the many-body interactions can be ignored.
Therefore an important question is, what happens when interactions
are turned on? Nature of collective excitations of the helical
metallic states~\cite{Zhang2010} and possible instabilities~\cite{babak}
are of particular interest. If the interactions are 
of short range nature and are strong enough to destroy the
metallic state of the surface, a Mott insulating state in the 
surface is expected, albeit with surface spinon bands~\cite{Balents}. 
When the interactions are not strong enough to destroy the
metallic nature of the surface states, the Coulomb interactions 
within a surface of a TI doped away from Dirac point 
gives rise to plasmonic and zero sound collective excitations in the
helical metallic state of the surface~\cite{Zhang2010}. Such collective excitations require
a Fermi surface and hence disappear in the undoped surface Dirac cone, $\mu=0$. 

The Dirac cone of undoped graphene supports a particle-hole {\em bound state} --
i.e. an split-off state below the free particle-hole continuum (PHC) -- 
which disperses linearly with momentum~\cite{BaskaranJafari,JafariJPCM,Ganesh}. 
In contrast to the anti-bonding nature of plasmon excitations~\cite{HwangSarma},
a minus sign required to render the interaction vertex attractive is 
provided by the {\em exchange} interactions~\cite{BaskaranJafari}. 
The Dirac cone describing the topological surface states features
a spin-momentum locking~\cite{zhang2009} which makes the physical
spin inaccessible for the exchange process. The question is then,
can there be a mechanism to give rise to a bound state below the
PHC of the free Dirac fermions topological insulators?
The physical spin of the electrons in helical metals is locked
to momentum and hence the spin-exchange scattering requires 
their momentum to be reversed. But such a back-scattering is
prohibited by the chiral nature of carriers. Therefore unlike graphene~\cite{BaskaranJafari},
the combination of time-reversal symmetry and the chiral nature of
helical states prevents taking advantage of the minus sign (attraction)
in the exchange Coulomb vertex in the spin-triplet particle-hole channel.
In this work we show that when the TI is cut into a thin film
the two surfaces can be described by a pseudo-spin $\hat\tau$ taking
on two values $\tau^z(\equiv\tau)=\pm 1$. Then {\em the $\tau$-exchange 
scattering part of the Coulomb interaction will provide attractive interaction 
between the positive-energy state from one surface and a negative-energy
state from the other surface} and hence leads to a bound state in the
particle-hole channel which is if singlet nature with respect to
spin, but can be considered of a triplet nature with respect to
pseudo-spin $\hat\tau$.  This is akin to excitons of 
a semiconductor, but with particle and hole states coming from 
two different surfaces $\tau=\pm$ surfaces of the TI slab. 
It can also be thought of as the "pseudo-spin" version of a 
collective mode in the Dirac cone of graphene~\cite{BaskaranJafari,JafariJPCM}.
Since the constituent particle and hole pairs come from two 
different planes, the center of mass of this collective mode will be 
in a bisecting plane parallel to the surface. Therefore in nanoscopic
thin films of TIs where inter-surface Coulomb interactions are substantial
in addition to the opposite-chirality Fermionic modes corresponding to 
single-particle helical states on the $\tau=\pm$ surfaces of a TI, 
there will be an additional two dimensional branch of collective excitations 
with spin-singlet bosonic character. 
%Therefore upon narrowing TIs a bosonic branch also emerges
%in the low-energy sector of the effective theory required to
%describe the TIs.

For an slab of TI with thickness $\ell$, the Coulomb energy scale 
is be given by $e^2/(\epsilon_d \ell)$ where
$\epsilon_d$ is the relative dielectric constant of the insulating
medium (the bulk states) separating the two surfaces. The separation $\ell$
serves to tune the strength of interactions for a fixed bandwidth of surface states. 
For nanoscopic separations on the scale of $\ell\sim 10$ nm, in one hand
the strength of the Coulomb interaction will become comparable to the 
bandwidth of the surface states~\cite{Bi2Se3}. On the other hand such separation is 
large enough to prevent the direct hybridization of the atomic
orbitals on the opposite surfaces, and hence the single-particle
part of the spectrum is protected from gap opening and will
continue to be a gapless Dirac theory.
Increasing the strength of $\tau$-off-diagonal interactions is expected to
give rise to fascinating possibilities, including the proposal of
topological excitonic condensation by Seradjeh and coworkers~\cite{babak}.

\section{Non-interacting Hamiltonian}
To fix the notations for the rest of the paper, let us summarize 
the non-interacting part of the Hamiltonian of a TI thin film
and discuss the relevant matrix elements. 
The low-energy effective theory describing the gapless
metallic states on the surface of a TI is the so called 
"helical metal" and is given by~\cite{zhang2009},
$
   H_0=v_F\sum_\vk \cdag_\vk \hat n.(\vk\times \vec\sigma)c_\vk
$
where the two-dimensional momentum $\vk$ (we set $\hbar=1$) is
in the surface characterized by normal vector $\hat n$. The momentum
is locked to its spin $\vec\sigma$. We use the spinor notation 
$\cd_\vk=(\cd_{\up\vk},\cd_{\down\vk})$ where $\up$ and $\down$
are the $z$-component of the spin of electrons. 
When two opposite surfaces labeled by the pseudospin values $\tau=\pm$ are brought 
to nano-meter distance $\ell$, as long as
the surfaces are not close enough to allow for the
overlap of atomic wave-functions the Hamiltonian 
of the thin film will be given by two copies of the
above Dirac model as,
\be
   H_0^{\rm slab}= v_F \sum_{\tau\vk} \cdag_{\tau\vk} 
   \left[\tau\hat z.(\vk\times \vec\sigma)\right] c_{\tau\vk}.
   \label{helical.eqn}
\ee

A constant energy configuration within this Hamiltonian is described 
by a circulating pattern of in-plane spin texture oriented perpendicular 
to their two dimensional momentum vector $\vk$~\cite{hasan}. If we denote the azimuthal
angle of $\vk$ with respect to the $k_x$ axis with $\theta_\vk$, then
in terms of the $\pi/2$ rotated momentum vector $\tilde k$ whose azimuthal angle
is $\varphi_\vk=\theta_\vk+\pi/2$, the $2\times2$ Hamiltonian matrix
in the above equation will be given by,
\be
   h_{\tau\vk}=\tau v_F k\left(
   \begin{array}{cc}
      0	& e^{-i\varphi_\vk}\\
      e^{i\varphi_\vk} & 0
   \end{array}
   \right),
\ee
where $k=|\vk|=\sqrt{k_x^2+k_y^2}$.
The following transformation from $c_{\tau\sigma\vk}$-basis 
to $a_{\tau\lambda\vk}-$basis defined by
\be
   \left(\begin{array}{cc}
   c_{\tau\up\bk}\\ c_{\tau\down\bk}
   \end{array}\right)
   =\frac{\tau}{\sqrt 2}
   \left(\begin{array}{cc}
   1 			& 	1\\
   e^{i\varphi_{\bk}}	& -e^{i\varphi_{\bk}}
   \end{array}\right)
   \left(\begin{array}{cc}
   a_{\tau,+,\bk}\\ a_{\tau,-,\bk}
   \end{array}\right),
\ee
brings $H_0$ to diagonal format:
\be
   H_0^{\rm slab}=\sum_{\tau\lambda\bk} \eps_{\tau\lambda\bk}~
   \ad_{\tau\lambda\bk}a_{\tau\lambda\bk},
\ee
where the subscript $\lambda=\pm $ stands for the helicity and
the cone like dispersion is given by 
$\eps_{\tau\lambda\bk}=\tau\lambda v_F k\equiv\tau\lambda\eps_\vk$.
The eigen mode $|\tau\lambda\vk\rangle$ of the helical metal
is created by
\be
   \ad_{\tau\lambda\vk}|\mbox{vac}\rangle =\frac{\tau}{\sqrt 2}
   \left(\cd_{\tau\up\vk}+\lambda e^{i\varphi_\vk} \cd_{\tau\down\vk} \right)
   |\mbox{vac}\rangle.
\ee
The above eigen-states correspond neither to 
$|\!\!\up\rangle$ nor $|\!\!\down\rangle$ spin states.
Therefore, while the energy eigen-modes correspond to spin-half 
($S^2|\tau\lambda\vk\rangle=3\hbar^2/4|\tau\lambda\vk\rangle$), 
their spin has no component perpendicular to the surface, i.e.
$\langle\vk\lambda|S_z|\vk\lambda\rangle=0$.
Rather at each $\vk$ point the eigen-state has a definite 
spin angular momentum in the transverse direction $\tilde k$ which is
in the plane of the surface and perpendicular to $\vk$. 
Such a locking between the
spin angular momentum on the linear momentum
roots in the coupling between spin and orbital motion
and forces the planar spin of the positive and negative energy 
states in the same plane to be always in a opposite directions~\cite{hasan,Bi2Se3}.
Therefore low-momentum transfer excitations between negative and
positive energy states will always correspond to spin-singlet
particle-hole pairs. So it is not possible to take advantage of
the spin-flip (exchange) processes in order to stabilize a 
bound state in the particle-hole channel as long as a fixed
surface is concerned. 

\begin{figure}[b]
  \begin{center}
  \includegraphics[width=4.0cm]{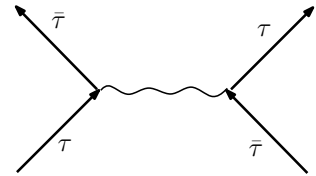}
  \caption{Basic $\tau$-exchange interaction responsible for
  the formation of inter-surface singlet exciton mode.
  The minus sign required to make the attraction interaction
  is generated by the exchange of $\tau$ attribute of the
  incoming states.
  }
  \label{diag00.fig}
  \end{center}
\end{figure}

However, when the other surface of the TI is brought closer,
the surfaces on the two sides of the slab can be labeled by $\tau=\pm$.
The index $\tau$ for an infinitely large plane will be an attribute of 
the electron motion (constant of motion) and therefore a {\em pseudo-spin} 
$\tau$ emerges very naturally. The Coulomb interaction in the channel that 
"exchanges" the $\tau$ attributes (Fig.~\ref{diag00.fig}) can take advantage 
of this pseudo-spin structure. Hence the Coulomb repulsion between 
electrons translates into attraction between a positive energy state 
in one plane (say $\tau=+$) and the negative energy state in the opposite 
plane ($\tau=-$) and hence is expected to bind them below the PHC. 

\section{Collective Excitations in Slab of Topological Insulator}
In the plasmon and zero sound excitations considered in~\cite{Zhang2010}
for a single value of $\tau$, only states with a fixed value of $\lambda$
are involved, as they are intra-band excitations. But when the
chemical potential is at the Dirac node, $\mu=0$, both positive
and negative energy states are expected to play role in the 
formation of the particle-hole bound state. 
The non-interacting part of the Hamiltonian $H_0^{\rm slab}$ 
describes electrons at $|\tau\lambda\vk\rangle$ states, where
the sign of energy is given by the product $\tau\lambda$ of the
pseudo-spin value and the helicity. Hence a positive energy state
from one plane and the negative energy state from the opposite plane
must have the same helicity $\lambda$.  For an electron of 
given energy traveling in some direction in the surface 
$\tau$ whose planar spin is $\sigma$, there will be another state at the
same energy corresponding to propagating electron in the opposite direction 
in plane $\bar\tau$ with the same planar spin $\sigma$. This can be alternatively 
thought of as a hole in the $\bar\tau$ surface whose spin is $\bar\sigma$ and
propagates in the same direction as the electron in state corresponding to 
$|\tau,\sigma\rangle$. The resulting attraction between the 
$|\tau,\sigma\rangle$ electron and $|\bar\tau,\bar\sigma\rangle$ hole
propagating in the same direction binds them in the spin-singlet
channel. 

Now let us put the above idea in a formal setting. To do this, 
we employ an interaction term of the following form:
\be
   H^{\rm int}=\frac{1}{2\cal A}\sum_{\vq\tau} \rho_{\vq\tau} \rho_{-\vq\bar\tau} v(\vq),
   \label{hint.eqn}
\ee
which is the standard density-density Coulomb interaction and the
density fluctuation $\rho_{\vq\tau}$ is the Fourier component of 
the density operator at surface $\tau$ and $\cal A$ is the area 
of the thin film under consideration. A peculiar feature of the
following problem would be due to the interaction $v(\vq)$: 
The momentum $\vq\equiv(q_x,q_y)$ is a two dimensional momentum and the Coulomb
interaction is between two planes separated by $\ell$ along the
$z$-direction. Hence $v(\vq)$ is a partial Fourier transform
or the three dimensional Coulomb potential with respect to 
planar momentum $\vq$, which will be given by,
\be
   v(\vq) = \frac{e^2}{4\pi\epsilon_d}\frac{\exp({-q\ell})}{q},
   \label{vq.eqn}
\ee
where the two surfaces are assumed to be at $z=0$ and $z=\ell$.
In the basis where the $H_0^{\rm slab}$ is diagonal,
the density operator at surface $\tau$ is represented as,
\be
   \rho_{\vq\tau} = \sum_{\lambda\lambda'\vk} \ad_{\tau\lambda'\vk'} 
   \langle \tau\lambda'\vk'|\tau\lambda\vk\rangle ~a_{\tau\lambda\vk},
\ee
where $\vk'=\vk+\vq$ is understood, and the wave-function overlap
factors are,
\be
   F_{\tau'\lambda'\vk',\tau\lambda\vk}\equiv  
   \langle \tau'\lambda'\vk'|\tau\lambda\vk\rangle 
   =\tau\tau'\frac{1+\lambda\lambda' e^{i\varphi_\vk-i\varphi_{\vk'}}}{2}.
\ee
In the $\tau=\tau'$ case the above overlap factor reduces to the familiar
factors in the two dimensional Dirac fermions of graphene~\cite{HwangSarma}.
In terms of the above density, the generic form of the interaction between
the two surfaces of the slab will be $v \ad_\tau \ad_{\bar\tau}a_{\bar\tau}a_\tau$.
This interaction in the $\tau$-flip channel depicted in Fig.~\ref{diag00.fig} will become,
$-v \ad_\tau \ad_{\bar\tau}a_\tau a_{\bar\tau}$. This means that the incoming
particles exchange their $\tau$ attribute, i.e. a particle from one surface
$\tau$ is scatters to the other surface $\bar\tau$, and at the same time
another particle scatters from surface $\bar\tau$ to surface $\tau$.  
This minus sign formally manifests in the following calculations as the $\tau\tau'=-1$ factor
multiplying the above wave-function overlap in the exchange channel.
The following analysis shows that this process is responsible for
formation of a exciton-like bound state between the positive energy
states of one surface and negative energy state of the opposite surface.
Let us consider a particle-hole excitation between the negative-energy
states of one surface and the positive energy states of the other
surface which will have the following form:
\be
   O_{\vK,\vQ}=a^\dagger_{++\vK'} a_{-+\vK},
\ee
where $\vK'=\vK+\vQ$ and hence the total momentum of the particle-hole 
pair is $\vQ$. Equation of motion for this operator is
$
   i\partial_t O_{\vK,\vQ}=[O_{\vK,\vQ},H_0+H^{\rm int} ].
$
The commutator of $O_{\vK,\vQ}$ with $H_0$ is,
\be
   \left[O_{\vK,\vQ},H_0 \right] = 
   -\left(\eps_\vK+\eps_{\vK+\vQ}\right) O_{\vK,\vQ},
   %\ad_{+,\vK+\vQ,+} a_{+,\vK,-}
   \label{h0EOM.eqn}
\ee
which upon identification of $i\partial_t\equiv -\omega\times$
means that as far as the the non-interacting part of the
Hamiltonian for the slab is considered the operator $O_{\vK,\vQ}$
creates a free particle-hole pair with energy $\eps_\vK+\eps_{\vK+\vQ}$.
The commutation relation with $H^{\rm int}$ after a Hartree 
decomposition to close the equation of motion gives~\cite{JafariJPCM,DemlerZhang},
\bearr
   &&\left[O_{\vK,\vQ},H^{\rm int}\right]\approx -\frac{v(\vQ)}{\cal A}
   \left[n_{++\vK'}-n_{-+\vK} \right]\times \nn\\
   &&\sum_{\vp\vp'} 
   F_{++\vp',-+\vp}
   ~a^\dagger_{++\vp'}a_{-+\vp} ~\delta_{\vp',\vp+\vQ}
   \label{h1EOM.eqn}
\eearr

Note that to obtain the above equation, the interactions in the
channel that exchanges the $\eta=\tau\lambda$ index of the incoming 
particles is essential. 
The equations of motions~\eqref{h0EOM.eqn} and~\eqref{h1EOM.eqn}
suggest that the collective operator defined in the last line of
Eq.~\eqref{h1EOM.eqn} satisfies the following collective mode
equation:
\be
  \frac{1}{v(\vQ)}=\frac{1}{\cal A}\sum_{\vK}\frac{n_{++\vK'}-n_{-+\vK}}
  {\omega-(\eps_{\vK}+\eps_{\vK'})}\times
  \frac{1+e^{i\theta_{\vK\vK'}}}{2}\equiv \chi_0,
  \label{mode.eqn}
\ee
where $\vK'=\vK+\vQ$ is understood. Moreover 
$\theta_{\vK\vK'}=\varphi_\vK-\varphi_{\vK'}$, and 
$n_{\tau\lambda\vK}$ is the expectation value 
$\langle a^\dagger_{\tau\lambda\vK}a_{\tau\lambda\vK}\rangle$.
Due to the symmetry under $\vQ\to -\vQ$ in the above equation,
the $\sin(\theta_{\vK\vK'})$ does not contribute to. 
The remaining integral is identical to those appearing in the
polarization of undoped graphene which has been calculated by
many authors~\cite{Son2007}, and gives rise to following simple form:
\be
   \chi_0 = \frac{Q^2}{16\sqrt{Q^2v_F^2-\omega^2}}.
\ee
This bubble is real and positive inside the gap below the particle-hole continuum $\omega < Qv_F$.
Hence the equation for the dispersion relation of the inter-surface spin-singlet exciton
will become,
\be
   \omega = Q v_F \left(1- \tilde\alpha^2 e^{-2Q\ell}\right)^{1/2},
   \label{boson.eqn}
\ee
where $\tilde\alpha=\frac{1}{64\pi}\frac{{\rm e}^2}{v_F}$ plays 
a role similar to fine structure constant.
Indeed it can be written as $\tilde\alpha=\frac{c}{v_F}\frac{\alpha}{64\pi}$
where $\alpha\approx 1/137$ is the fine structure constant. For 
Fermi velocities in the scale of $10^4-10^5$m/s in the Bismuth family~\cite{Allen} 
one expects the range of values $\tilde\alpha\sim 10^{-2}-10^{-1}$.
According to this expression, an instability in the $\vQ\to 0$ limit
can be triggered when $\tilde\alpha$ can become large which is only
possible by finding materials with smaller $v_F$.
For the Bismuth family TIs, given the very small value of $\tilde\alpha$, 
such instability does not seem feasible. Smallness of $\tilde\alpha$ allows
us to Taylor expand the RHS of the dispersion relation to get,
\be
   \omega \approx Q v_F \left(1- \frac{1}{2}\tilde\alpha^2 e^{-2Q\ell}\right),
   \label{boson2.eqn}
\ee
which defines the binding energy of the bosonic mode
with respect to the lower edge of free particle-hole continuum as,
\be
E_B=Q v_F \frac{\tilde\alpha^2}{2} e^{-2Q\ell}.
\ee
If the thickness $\ell$ of the slab becomes very large, 
the exponential factor will diminish the binding energy
and the bosonic mode will be absorbed to the lower 
edge of the continuum of free particle-hole pairs. 
However when the $\ell$ is not very large, 
the dispersion relation~\eqref{boson2.eqn} defines
a relativistic branch of bosons. The boson velocity
acquires the following form
\be
\frac{v_B}{v_F}=1+ \left(Q\ell-\frac{1}{2}\right) \ell\tilde\alpha^2 e^{-2Q\ell}
\ee
The maximum of the binding energy is obtained for 
$2Q\ell=1$ and is $E_B^{\rm max}=\tilde\alpha^2 v_F/(4\ell)$.
Below this point $v_B<v_F$ and above it $v_B> v_F$.
At this point itself, $v_B$ becomes equal to $v_F$.
Therefore for practical purposes in the low momentum region 
define by $2Q\ell \lesssim 1$, one may assume that the bosonic branch 
is almost degenerate with Fermionic branch and has the
approximate dispersion relation
\be
   \omega= Q v_B,~~~~~~v_B\lesssim v_F.
\ee
For $\ell=1$nm, we expect $E_B^{\rm max}\sim 10^{-2}$ meV.
Although the direct observation of the above excitonic
mode is difficult, but it may have indirect consequences
on the spectral properties of the Fermions themselves.

\section{Summary and conclusions}
Using random phase approximation implemented by the equations
of motion we explicitly constructed spin-singlet $\hat\tau$-triplet
collective operators for a nano-meter thick film of topological insulator.
The $\tau$-exchange part of the Coulomb interaction gave rise
to an attraction between the positive energy states of one surface
and the negative energy states of the other surface. Such attraction
results in a binding bosonic mode below the continuum of free
particle-hole excitations. The binding energy has a characteristic
$\exp(-Q\ell)$ dependence which vanishes when the thickness $\ell$ of the
slab becomes very large and hence reduced to proper limit. Although 
the binding energy of the present inter-surface excitonic mode is 
a fraction of meV, but given its chiral nature, it is expected
to leave Kerr and Faraday signatures in optical absorption~\cite{Lozovik}.
The self-energy corrections from such inter-surface excitons will
indirectly show up in angular resolved photo-emission spectroscopy 
and scanning tunneling spectroscopy. Such effects can become
more pronounced by approaching the excitonic instability~\cite{Polini}.

\section{acknowledgments}
This research was completed while the author was
visiting Yukawa Institute for Theoretical Physics
by the fellowship S13135 from Japan Society for Promotion of Science.
Useful discussions with T. Tohyama, N. Kawakami, H. Fukuyama, M. Ogata,
Y. Fuseya, A. Mishchenko, Bohm-Jung Yang and M. S. Bahramy is acknowledged. 
We thank W. Koshibae for hospitality during a short visit to RIKEN.

\end{document}